\documentstyle[12pt]{article}	
\textheight 9in  \topmargin -.5in   \def\baselinestretch{1.2}
\def\Eq{\begin{equation}}	\def\End{\end{equation}}
\def\Eqa{\begin{eqnarray}}	\def\Enda{\end{eqnarray}}
\def\Endl#1{\label{#1} \End}	\def\Endla#1{\label{#1} \Enda}
\def\puteq#1{eq.~(\ref{#1})}	\def\Puteq#1{Eq.~(\ref{#1})}

\def\lapprox{\mathrel{\scriptstyle{\buildrel < \over \sim}}}
\def\pmb#1{\setbox0=\hbox{#1}\kern-.01em\copy0\kern-\wd0
  \kern.02em\copy0\kern-\wd0\kern-.01em\raise.017em\box0}
\def\ord#1{{\cal O}(#1)}  \def\d{{\rm d}}  
\def\mzs{m_0^2}  \def\mzsc{m_{0C}^2}  \def\lz{\lambda_0}
\def\L{\Lambda}  \def\lL{\lambda_L}  \def\k{\kappa}  \def\d{{\rm d}}
\def\LaC#1{(\L a)_{C#1}}  \def\eg{{\it e.g.}}
\def\pff{$\lambda\phi^4_{\;4}$}  \def\pst{$\lambda\phi^6_{\;3}$}
\def\pft{$\lambda\phi^4_{\;3}$}  \def\0{\phantom{0}}

\newcounter{xpos} \newcounter{ypos}
\def\Bp{\begin{picture}}	\def\Ep{\end{picture}}
\def\core#1#2#3#4#5#6{\raisebox{#1pt}{\Bp(#2,#3) \setcounter{xpos}{#4}
  \setcounter{ypos}{#5} #6 \Ep}}
\def\Mu{\addtocounter{ypos}{ 20}}
\def\Ca{\put(\value{xpos},\value{ypos}){\circle{20}}}		
\def\Cai{\Ca \put(\value{xpos},\value{ypos}){\circle{16}}}	

\input epsf
\def\boxit#1{\vbox{\hrule\hbox{\vrule{#1}\vrule}\hrule}}
\def\INSERTFIG#1#2#3{\epsfxsize=#1in \centering
  \vbox{\hfil\boxit{\epsffile{#2}}\hfill} {\sl #3}\bigskip}

\begin{document}
\begin{titlepage}
\begin{center}
March, 1994	\hfill CMU-HEP94-10\\
{}~		\hfill DOE/ER/40682-64\\
{}~		\hfill hep-lat/9403021\\
\vskip1.5in {\Large\bf The Lattice Cutoff for \pmb{\pff}\ and \pmb{\pst}}
\vskip.3in
{\bf David E. Brahm}\footnote{Email: \tt brahm@fermi.phys.cmu.edu}\\
  \vskip.2cm {\it Carnegie Mellon Physics Dept., Pittsburgh PA 15213}
\end{center}

\vskip.5in
\begin{abstract}
We analyze the critical line of \pff\ perturbatively in the bare coupling
$\lz$, by setting the daisy-improved renormalized mass to zero.  By comparing
to lattice data, we can then quantify the relation between the continuum cutoff
and the lattice spacing; for the 4-dimensional hypercubic lattice we find
$\LaC4 = 4.893$.  We perform a similar analysis for \pst, and find in 3
dimensions $\LaC3 = 4.67$.  We present two theoretical predictions for $(\L
a)$.  For small $\lz$, both the critical line and the renormalized mass near
criticality are easily and accurately calculated from the lattice input
parameters.
\end{abstract}
\end{titlepage}
\setcounter{footnote}{0}


\section{Introduction: The Lattice Cutoff}

Placing a field theory on a lattice (of physical spacing $a$) provides an
effective cutoff $\L \sim 1/a$, and breaks its translational and rotational
invariance.  The invariance is restored in the physics near a phase transition.
By analyzing the critical line of a simple scalar theory (with quadratic
divergences), we can find the numerical value of $(\L a)$ for that theory.

Contrary to the usual interpretation of the lattice phase transition (in which
the renormalized mass $m^2$ is considered fixed while $a\to0$), we are taking
$a$ to represent a fixed physical distance (or scale $\L$), while $m^2$ is
tuned to zero (by varying the bare mass $\mzs$, which is a lattice input
parameter).  ``Triviality'' merely states that the renormalized $\lambda\to0$
at the same time, which does not matter to us since our analysis uses only the
bare coupling $\lz$.

Once $(\L a)$ is known, then even away from the critical point the renormalized
mass $m^2$ (at scale $\mu=m$) can be easily calculated from the bare mass
$\mzs$ [see \puteq{renms}].  This will prove useful when the system is placed
in a thermal bath (implemented by squeezing the lattice in the time direction),
and the symmetry that was broken at $T\!=\!0$ is restored at some critical
$T_c$.  We will then know $T_c$ as a function of the renormalized mass.

The critical line of \pff\ has been analyzed before, notably by L\"uscher and
Weisz\cite{kexp,lang}, and we will treat their results as ``data''.  Our
method, by comparison, only works for small bare coupling ($\lz\ll 4\pi^2$),
but is much simpler, providing comparable accuracy (in the region of validity)
from a calculation of only two diagrams with massless propagators.

\section{\pmb{$\lambda\phi^4$}\ Theory on a Lattice}

The action of a 1-component scalar field in 4 Euclidean spacetime dimensions
\Eq S = \int {\rm d}^4 x \left[ {1\over2}(\partial\phi)^2 + {\lz\over4}
  \phi^4 - {\mzs\over2} \phi^2 \right]  \Endl{Scont}
can be put on an $L^4$ hypercubic lattice:
\Eq S_{latt} = \sum_x \left[-2\k \sum_{\mu=1}^4 \Phi_x \Phi_{x+\mu} +
  \Phi_x^2 + \lL (\Phi_x^2-1)^2 \right]  \End
by the change of variables
\Eq \Phi = {\phi a \over \sqrt{2\k}}, \qquad \lL = \k^2 \lz, \qquad
  \k = {1 - 2\lL \over 8 - \mzs a^2}  \End
The ``bare'' couplings $\lz$ and $\mzs$ are defined at a renormalization scale
(cutoff) $\L \sim 1/a$.  We will hereafter usually set $a=1$.

\subsection{The Critical Line \pmb{$\mzsc(\lz)$ } from Perturbation Theory}

The scalar theory at a lower scale $\mu$ is approximately given by
\puteq{Scont} but with $\lz\to\lambda$, $\mzs\to m^2$.  More precisely, $-m^2 =
V''_{\rm eff}(0)$ and $\phi_{\rm min}^2 = m^2/\lambda$, where $V_{\rm eff}$ is
the effective potential at scale $\mu$ and $V'_{\rm eff}(\phi_{\rm min})=0$.
To 2-loop order,
\Eq \Bp(32,15) \put(0,0){\line(1,0){32}} \put(0,3){\line(1,0){32}}
    \put(4,7){$-m^2$} \Ep \;=\;
  \Bp(32,15) \put(0,1){\line(1,0){32}} \put(4,7){$-\mzs$} \Ep \;+\;
  \core{0}{32}{20}{16}{10}{\Cai \put(0,0){\line(1,0){32}}} \;+\;
  \core{-10}{32}{20}{16}{10}{\put(0,10){\line(1,0){6}} \Cai
    \put(26,10){\line(1,0){6}} \put(6,9){\line(1,0){20}}
    \put(6,11){\line(1,0){20}}} \;+\;\cdots
  \Endl{2loop}
We have dressed propagators, but not vertices\cite{way}, to sum daisy-type
diagrams; thus the two-scoop diagram is not included because it is contained in
the dressed one-scoop diagram\cite{itz}.  The second-order phase transition
occurs when $m^2\to0$:
\Eq \mzsc = \L^2 \left[ {3\lz\over16\pi^2} - {3\lz^2\over64\pi^4} + {z_3\lz^3
  \over256\pi^6} + \ord{\lz^4} \right] \Endl{2lpred}
where we have included an unknown third-order coefficient $z_3$.  We fit
lattice ``data'' for $\k(\lL)$ from \cite{kexp} to \puteq{2lpred}.
Specifically, the first 13 points ($\lz\lapprox 10$) from their Table 1 yield
\Eq \LaC4 = 4.893\pm 0.003, \qquad z_3 = 3.03\pm 0.06 \Endl{LaC4}
With this value of $\L$ \cite{agodi,warn} we have plotted in Fig.~1 the 1-loop
(dashed) and 2-loop (solid) predictions from \puteq{2lpred}, along with the
first 11 data points.

\subsection{4-Component \pmb{$\lambda\phi^4$}}

A similar analysis for 4-component \pff\ theory gives, in place of
\puteq{2lpred},
\Eq \mzsc = \L^2 \left[ {3\lz\over8\pi^2} - {3\lz^2\over32\pi^4} + {z'_3\lz^3
  \over128\pi^6} + \ord{\lz^4} \right] \Endl{c2lpred}
The first 15 points ($\lz\lapprox 12$) from ref.~\cite{kexp4}, Table 1, yield
\Eq \LaC4 = 4.905\pm 0.004, \qquad z'_3 = 4.20\pm 0.05 \Endl{cLaC4}
Using this $\L$, we have plotted in Fig.~2 the 1-loop (dashed) and 2-loop
(solid) predictions from \puteq{c2lpred}, along with the first 12 data points.
$\LaC4$ agrees well with \puteq{LaC4}.

\subsection{Away from Criticality}

\Puteq{2loop} can be used away from the critical line as well.  For small
positive $(\mzs-\mzsc)$ the effect of a non-zero renormalized mass $m$ running
around the loops is negligible, and
\Eq {1\over\chi} = V''_{\rm eff}(\phi_{\rm min}) = 2m^2 \approx 2(\mzs-\mzsc)
  \Endl{1ochi}
The susceptibility $\chi = L^4 (\langle\bar\phi^2\rangle - \langle|\bar\phi|
\rangle^2)$ is measured directly on the lattice.  In Fig.~3 we plot $1/\chi$
vs.\ $\mzs$ for 1-component \pff\ [\puteq{Scont}] with $\lz=0.5$ on an
$8^4$ lattice, and we see the linear fit is quite good (up to about $\mzs=0.5$,
in fact).  Using lattices from $L=4$ to $L=8$ we found
\Eq {1\over\chi} = \left[ (2.04 \pm 0.05) + {(6 \pm 3)\over L^2} \right]
  \left\{ \mzs - \left[ (0.224 \pm 0.001) + {(1.00 \pm 0.05)\over L^2} \right]
  \right\} \End
The slope is consistent with 2 [\puteq{1ochi}], and $\mzsc=0.224$ agrees with
ref.~\cite{kexp} and \puteq{2lpred}.

\section{\pmb{$\lambda\phi^6$}\ Theory in 3 Dimensions}

To determine $\LaC3$ we analyzed the theory\cite{pstr}
\Eq S = \int {\rm d}^3 x \left[ {1\over2}(\partial\phi)^2 + {\lz\over6}
  \phi^6 - {\mzs\over2} \phi^2 \right]  \Endl{S63}
We chose this theory because $\lz$ is dimensionless, and so $\mzs/\L^2$ is
again just a polynomial in $\lz$.  To 2 loops,
\Eq \Bp(32,15) \put(0,0){\line(1,0){32}} \put(0,3){\line(1,0){32}}
    \put(4,7){$-m^2$} \Ep \;=\;
  \Bp(32,15) \put(0,1){\line(1,0){32}} \put(4,7){$-\mzs$} \Ep \;+\;
  \core{-19}{32}{40}{16}{10}{\Cai\Mu\Cai \put(0,20){\line(1,0){32}}}
  \Endl{2lp63}
so the phase transition occurs at:
\Eq \mzsc = \L^2 \left[ {15\lz\over4\pi^4} + z_2 \lz^2 + \ord{\lz^3} \right]
  \Endl{ptg}
We could not calculate $z_2$ because infrared divergences arise in a 4-loop
diagram, though the data will show that $z_2$ is finite.

We performed lattice simulations for $L=5$ to $L=10$, using the standard
Metropolis algorithm\cite{metro,lang}, and fit the broken-phase
susceptibility to
\Eq {1\over\chi} = \left[ A + {B\over L^2} \right] \left\{ \mzs - \left[ \mzsc
  + {C\over L^2} \right] \right\} \Endl{cfit}
The results are in Table~1, and yield
\Eq \mzsc = (0.84\pm 0.01)\,\lz - (0.23\pm 0.03)\,\lz^2 \End
Compared to \puteq{ptg}, this gives
\Eq \LaC3 = 4.67\pm 0.03, \qquad z_2 = 0.011\pm 0.001 \Endl{LaC3}

\begin{table} \centering \begin{tabular}{|r|cccc|} \hline
  $\lz$ & $A$ & $B$ & $\mzsc$ & $C$ \\ \hline
  $0.10$ & $2.19\pm 0.04$ & $ 37\pm 2$ & $0.0792\pm 0.0012$ & $2.19\pm 0.07$ \\
  $0.15$ & $2.39\pm 0.02$ & $ 21\pm 1$ & $0.1220\pm 0.0012$ & $2.08\pm 0.07$ \\
  $0.20$ & $2.38\pm 0.04$ & $ 18\pm 3$ & $0.1596\pm 0.0016$ & $2.13\pm 0.11$ \\
  $0.30$ & $2.39\pm 0.04$ & $ 14\pm 2$ & $0.2313\pm 0.0031$ & $2.07\pm 0.19$ \\
  $0.40$ & $2.45\pm 0.02$ & $\09\pm 1$ & $0.2988\pm 0.0017$ & $2.08\pm 0.09$ \\
  \hline \end{tabular} \caption{$1/\chi$ for \pst.} \end{table}

\section{Theoretical Predictions of \pmb{$(\L a)$}}

\subsection{A Geometrical Prediction}

We now present a geometrical prediction for the value of $(\L a)$ on a
$D\!=\!4$ hypercubic lattice.  Wavevectors along a major lattice axis might see
a cutoff $2\pi/a$, but in other directions the distance across the hypercubic
cell is $r>a$.  Invoking rotational invariance, we calculate $\L$ as the
average over all directions of $\langle 2\pi/r \rangle$ (see Fig.~4):
\Eqa \LaC4 &=& 2\pi {2^4 \, 4!\over 2\pi^2} \int_0^{\pi/4} \d\phi
   \int_0^{\cot^{-1}(\cos\phi)} \sin\theta \,\d\theta
   \int_0^{\cot^{-1}(\cos\theta)} \sin^2\psi\cos\psi \,\d\psi \nonumber\\
  &=& {32\sqrt2\over\pi} \,\sin^{-1}(1/3) = 4.8954  \Endla{La}
which agrees well with the results \puteq{LaC4} and \puteq{cLaC4}.

This prediction generalizes immediately to other numbers of dimensions:
\Eq (\L a)_{C1} = 2\pi = 6.283, \qquad (\L a)_{C2} = 4\sqrt2 = 5.657, \qquad
  \LaC3 = 6\sqrt2 \,\cot^{-1}(\sqrt2) = 5.223 \End
Sadly, the predicted $\LaC3$ does not agree with \puteq{LaC3}.

\subsection{A Lattice Momentum Vector Prediction}

Following ref.~\cite{palma}, the 1-loop diagram in \puteq{2loop} can be
calculated as a sum over all allowed lattice momentum vectors (except $\hat
k=0$):
\Eq 3\lz \int^\L {\d^4 k\over(2\pi)^4} {1\over k^2} = {3\lz\L^2 \over 16\pi^2}
  \to {3\lz\over L^4} \sum_{\hat k} {1\over \hat k^2}, \qquad
  \hat k_\mu \equiv 2 \sin\left( \pi n_\mu \over L \right), \quad
  n_\mu \in \left( {-L\over2}, {L\over2}\, \right] \End
In the infinite volume limit this turns back into an integral
\Eq {3\lz\L^2 \over 16\pi^2} = 3\lz \int^\pi {\d^4 k\over(2\pi)^4}
  \left[\sum_\mu 4 \sin^2 (k_\mu/2) \right]^{-1} \End
Performing this integral (numerically) we find $\LaC4 = 4.946$, and in 3
dimensions we find analogously $\LaC3 = 4.989$.  Agreement with our
empirical results is reasonable (1\%) in $D\!=\!4$, but again poor in
$D\!=\!3$.

\section{Results and Conclusions}

Comparing the renormalization equation \puteq{2loop} to lattice ``data'' for
the \pff\ theory gives $\LaC4 = 4.893$ (or $4.905$ for the 4-component
version), in good agreement with our geometrical prediction (and fair agreement
with the lattice momentum vector prediction).  Similar analysis of \pst\ gives
$\LaC3 = 4.67$; here both theoretical predictions fail.  It is possible that
the infrared divergence at 4 loops is affecting the $\ord{\lz}$ result.  We are
thus uncertain whether the geometrical prediction's success with \pff\ is
coincidental.

In addition to the work described above, we performed a few simulations of
\pft\ on a cubic lattice, for which the 1-loop diagram of \puteq{2loop}
predicts $\mzsc = 3\lz\L/2\pi^2$, and found for that system $\LaC3 = 4.3
\pm 0.1$.  We also attempted an analysis of 4-component \pff\ on the F4
lattice, but (somewhat disturbingly) the data in \cite{F4} did not fit
\puteq{2lpred}.

Given the value of $\LaC4$, we can successfully predict the critical line of
\pff\ theory to $\ord{\lz^2}$ from a very simple two-loop calculation, in which
daisy-improvement of the diagrams makes the propagators massless while leaving
the couplings bare.  Conversely, the data suggests the 3-loop result should be
$3\lz^3/256\pi^6$ [\puteq{LaC4}].  Even away from criticality, this method
makes the calculation of the renormalized mass from the lattice input
parameters simple [\puteq{1ochi}].

Near the critical point, the renormalized mass of the $N\!=\!1\,$ \pff\ theory
on a $D\!=\!4$ hypercubic lattice (as $L\to\infty$) is
\Eq m^2 \approx \mzs - (4.893)^2 \left[ {3\lz\over16\pi^2} -
  {3\lz^2\over64\pi^4} \right] \Endl{renms}
We expect to be able to use this result in an upcoming lattice study of the
\pff\ theory at finite temperature.  By choosing $\mzs$ a bit larger than
$\mzsc$, and then squeezing the lattice in the time direction until symmetry is
restored, we will be able to map out $T_c(m)$.

\vskip.4in
The author wishes to thank Rajan Gupta, Richard Holman, Stephen Hsu, Christian
B. Lang, and Luis Lavoura for valuable discussions.  This work was supported
in part by the U.S. Dept. of Energy under Contract DE-FG02-91-ER40682.


\def\np#1{{\it Nucl. Phys.\ }{\bf #1}}
\def\pr#1{{\it Phys. Rev.\ }{\bf #1}}
\def\rb#1{{\it Rev.\ Bras.\ Fis.\ }{\bf #1}}
\def\jc#1{{\it J.\ Chem.\ Phys.\ }{\bf #1}}
\def\zp#1{{\it Z.\ Phys.\ }{\bf #1}}

\newpage
\def\baselinestretch{.8}


\newpage
\INSERTFIG{4.9}{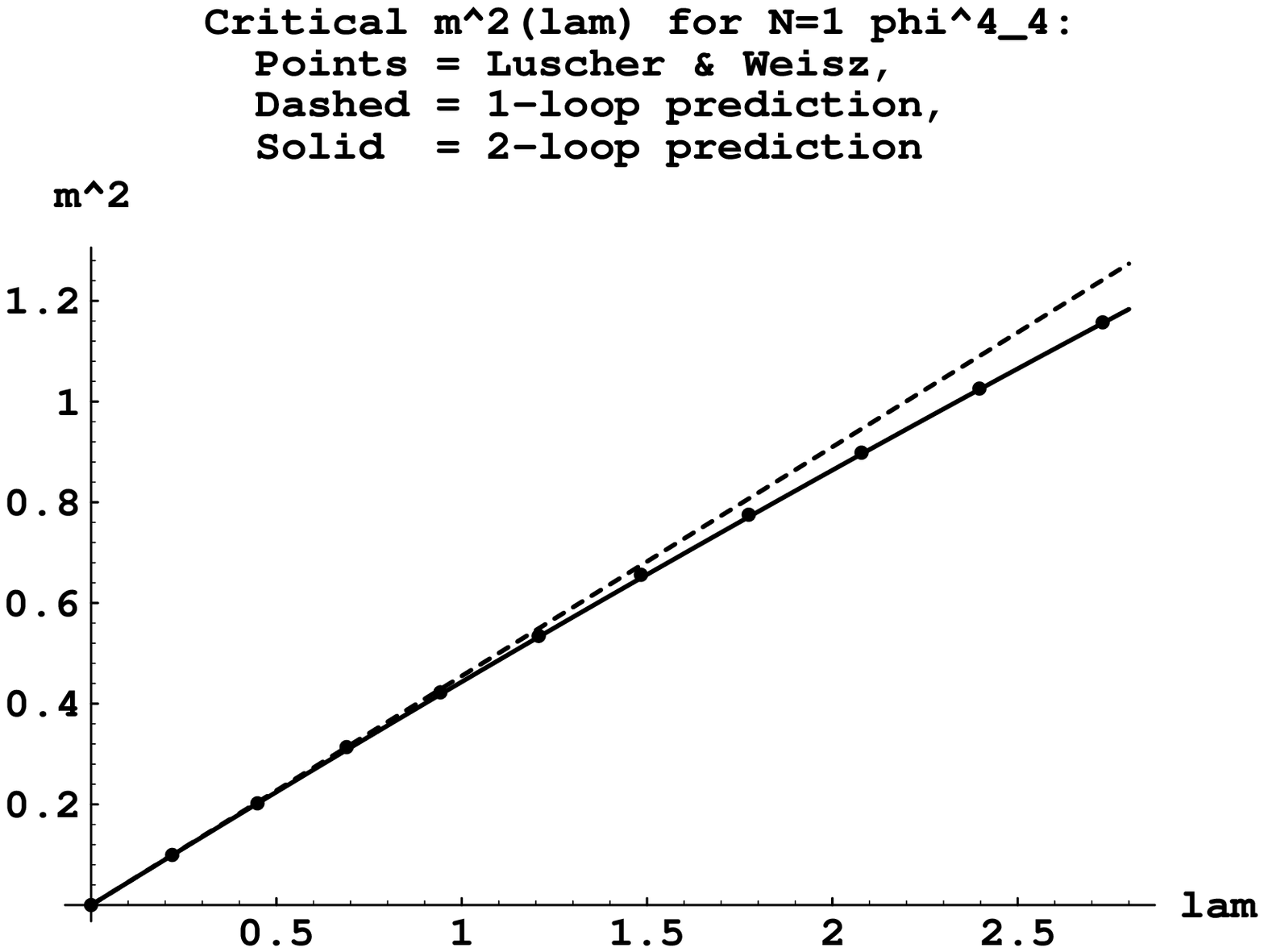}{Fig.~1: $N\!=\!1$ \pff\ critical line in the
  $\{\lz,\mzs\}$ plane: ``data'' and predictions.}
\vskip.2in
\INSERTFIG{4.9}{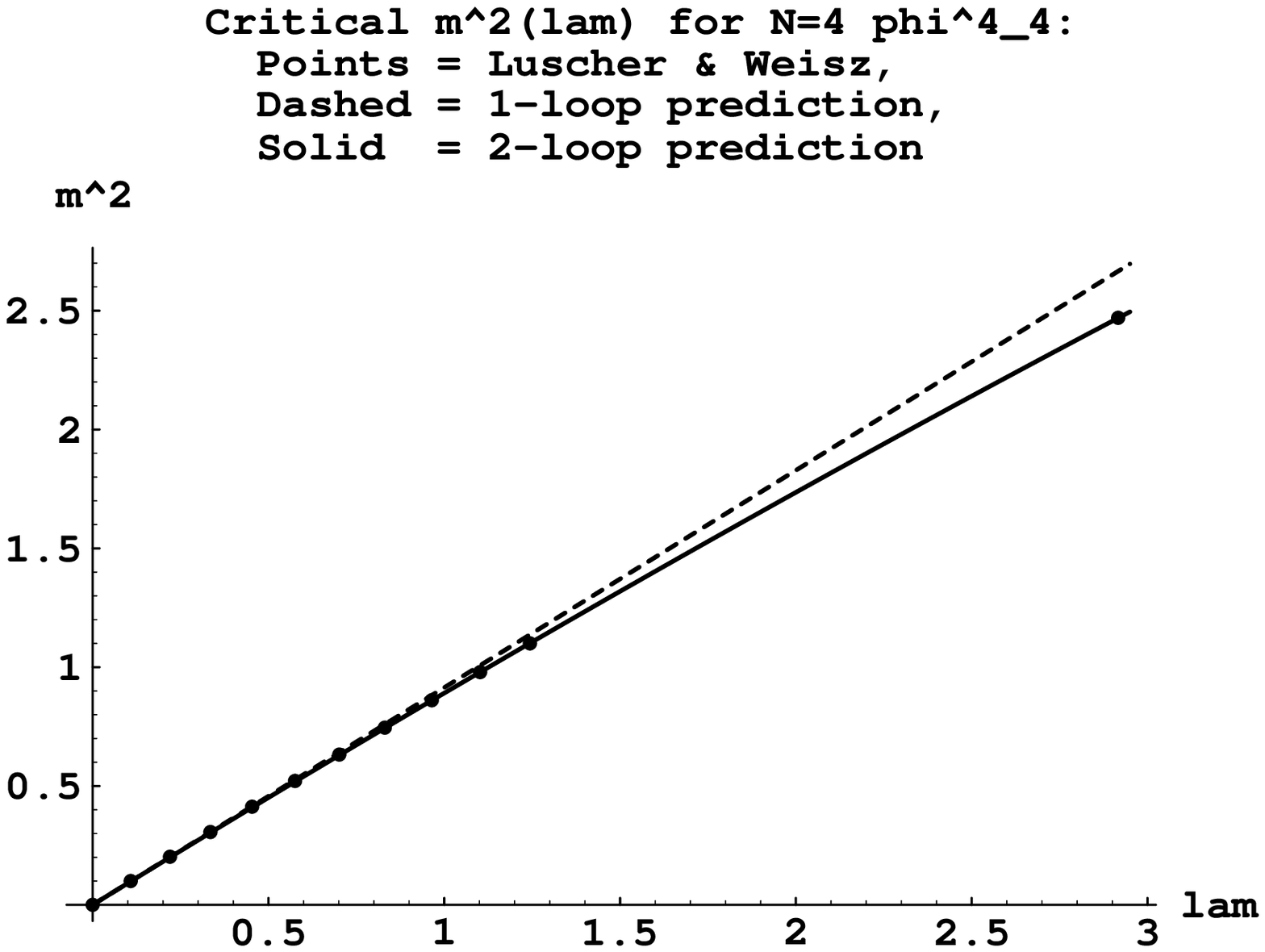}{Fig.~2: $N\!=\!4$ \pff\ critical line in the
  $\{\lz,\mzs\}$ plane: ``data'' and predictions.}
\vskip.2in
\INSERTFIG{5.1}{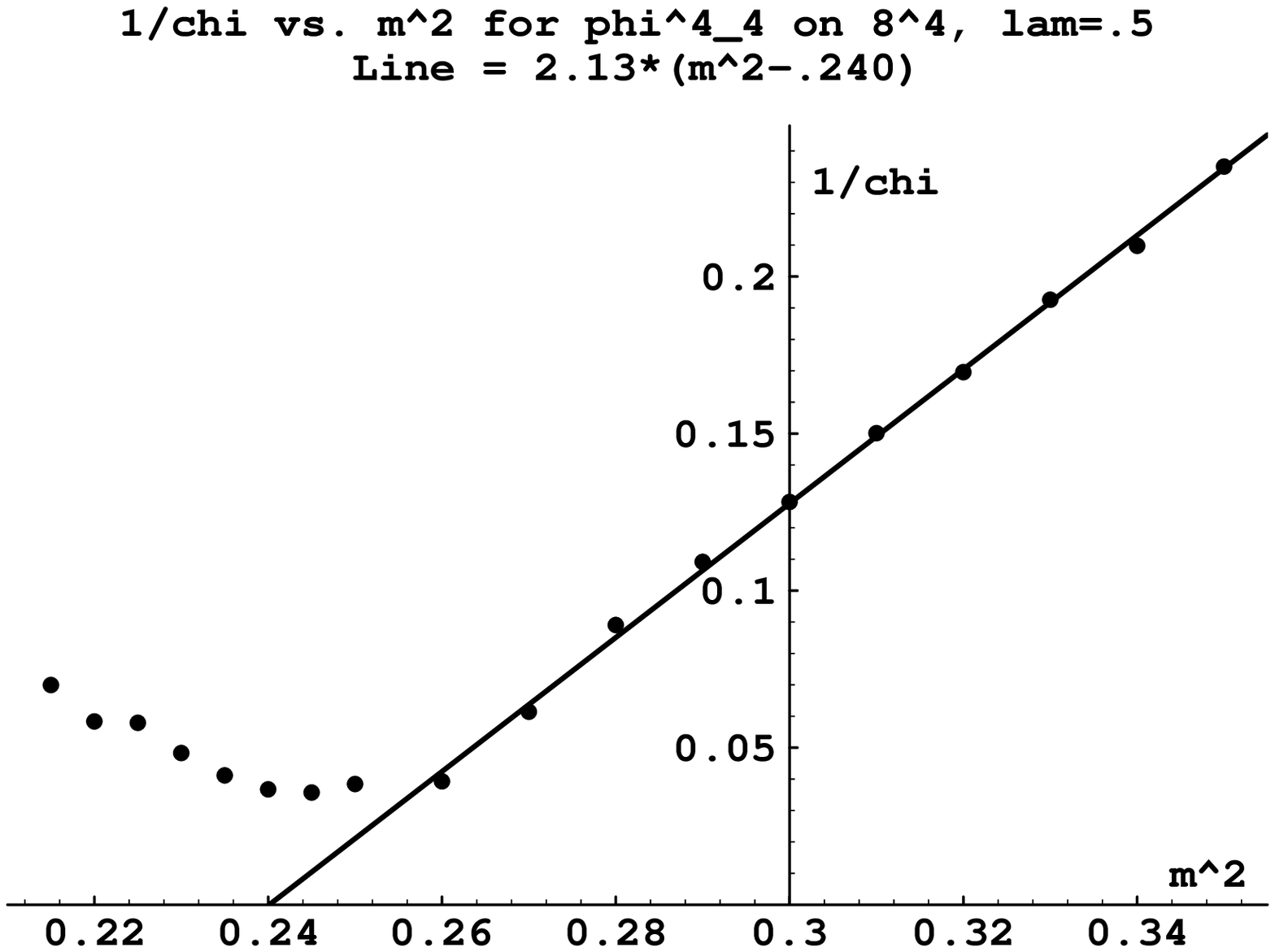}{Fig.~3: $1/\chi$ vs.\ $\mzs$ for 1-component \pff\
  with $\lz=0.5$ on an $8^4$ lattice.}
\vskip.2in
\def\boxit#1{#1}
\INSERTFIG{3.5}{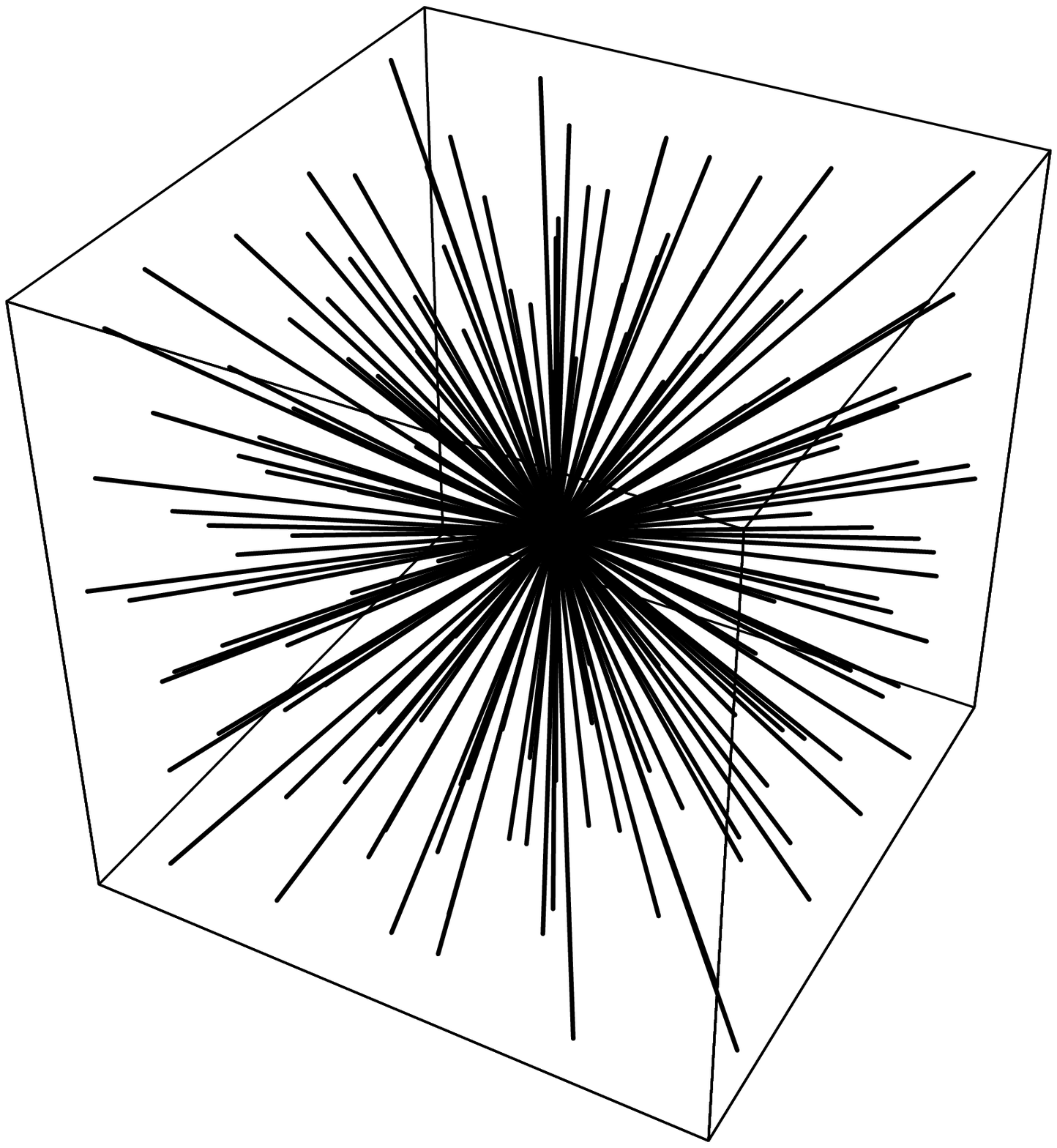}{Fig.~4: $\L$ is the average of $\langle 2\pi/r
  \rangle$, shown here in 3 dimensions.}


\begin{thebibliography}{9}
\bibitem{warn}	For the astute reader who sees a possible pattern emerging in
		  the coefficient $z_3\approx 3$, the pattern does not appear
		  to continue, sorry.
\end{thebibliography}
\end{document}